# Cooperative and collective effects in light of the maximum energy dissipation principle

Adam Moroz

Email: adasmaroz@icloud.com

(Article submitted to Physics Letter A, and published PLA 374 (2010) 2005-2010)

## Abstract

We compare the collective phenomena in physics and cooperative phenomena in biology/chemistry in terms of the variational description. The maximum energy dissipation employed and the cost-like functional was chosen according to an optimal control based formulation. Using this approach, the variational outline has been considered for non-equilibrium thermodynamic conditions. The differences between the application of the proposed approach to the description of cooperative phenomena in chemical/biochemical kinetics and the Landau free energy approach to collective phenomena in physics have been investigated.

## 1. Introduction

Identifying the similarity in the kinetic manifestation of collective effects in physics and cooperative processes in chemistry/biology can play a significant role in developing a better understanding of the nature of these phenomena. It is mostly interesting due to the difference in the types, scales, characteristic times and participants of the processes underlying the phenomena in these two different fields. It is also needed to take into account that these phenomena are nonlinear, moreover, extreme – and where they are involved in the regulatory mechanisms (biology), they provide the optimal way of control, including energetical effectiveness. Consequently, it is exceedingly interesting to study and compare these two types of processes from a phenomenological perspective, closely related to thermodynamics.

A fundamental extreme thermodynamic principle that has recently attracted substantial attention is the maximum entropy production (MEP) hypothesis, which many authors have formulated in different ways [1-6]. The MEP principle is closely related to the maximum energy dissipation (MED) principle [7-12]. Recently, some suggestions have been made regarding the maximum energy dissipation principle as directly related to the least action principle [11,12], together with its variational formulation for chemical kinetics.

However, the formulation of the maximum energy dissipation principle in terms of variational formalism still needs further illumination and study, especially in applied fields, like for example mentioned nonlinear similarities in physical kinetics and chemical and biochemical kinetics. In this Letter we present a comparison between the variational description of macroscopic collective effects in physics (based on the Landau free energy), and the cooperative effects in chemical and biochemical kinetics. This comparison is based on our former phenomenological outline [11,12] employed optimal control and based on the maximum energy dissipation principle.

## 2. Preliminaries

We can generalise the main points of results in [11,12] where the variational approach employing the MED principle for chemical thermodynamics was presented. In that outline, the classical mechanics was considered, where the control variables appeared as artificial variables in the equation $\dot{q} = u$, and therefore the N-dimensional vector of generalized velocities $\dot{q}$ become formally the control variable $u$. Then corresponding to classic mechanical dynamic Lagrange optimal control (OC) problem [11] becomes

$$S = \int_{t_1}^{t_2} \Lambda(u,q)dt = \int_{t_1}^{t_2} (T(u) - U(q))dt \to extr \qquad (1)$$

subject to $\dot{q} = u$. Here $S$ is the action, $\Lambda$ is the Lagrangian, $q$ is vector of generalized coordinates, $U$ is the potential term, $T$ is the kinetic term, $t$ is time, $(t_1,t_2)$ is the time interval, the control $u$ should be considered as having no restrictions. The technique to solve this problem is known as the Pontryagin maximum principle [13]. At the same time, as noticed in [11], from the cost-like perspective of optimal control, it is rather difficult to interpret the negative sign of the potential term in the Lagrangian Eq.(1) formulating the OC mechanical problem. When it has a positive sign, it is simple to interpret it from the OC as the penalty for not being in equilibrium. Also, it could be noted that the control $u$ has no real physical meaning.

In contrast to classical mechanics, in a nonlinear dynamical system case, we can write even more complicated relationship for constraints

$$\dot{\xi} = f(\xi, u-k), \quad \xi(t_0) = \xi_0, \qquad (2)$$

where $\xi$ is the extend from the equilibrium vector, $k$ is the vector of rate constants, $u$ – control vector . We can consider Eq. (2) as the dynamic constraints, that have much more complicated relationship than linear $\dot{q} = u$. Then we can also formally define an energetic cost (energetic loss for the regulation in case of metabolic network) for the rate constant deviation from an optimal value $k$, $\Phi = \Phi(u-k)$. The positive definite potential term we can write more specifically for dissipative kinetics as free energy $\Psi = \Psi(\xi)$. We also need take into account that the relaxational, dissipative processes are processes with no fixed final time – we cannot *a priori* suggest anything about time when the system will achieve the equilibrium $\xi=0$. Such a formulation corresponds to the open-end Lagrange problem. Taking it into account, we can write the minimisation functional following [11,12] as

$$S=\int_{t_0}^{\tau}\Lambda(\xi,u-k)dt=\int_{t_0}^{\tau}(\Phi(u-k)+\Psi(\xi))dt\to\min$$
$$\xi(t_0)=\xi_0 \qquad (3)$$
subject to $\dot{\xi}=f(\xi,u-k)$,

here with a fixed initial time $t_0$, unspecified final time $\tau$, fixed target state $\xi=0$. We will consider this OC problem as having no formal restrictions on the control variables $u$ and on the state variables $\xi$. According to the Pontryagin maximum principle, we can write the OC Hamiltonian as

$$H(\xi,u,p)=-\Phi(u-k)-\Psi(\xi)+pf\,. \qquad (4)$$

Since the final time $\tau$ is free and all $\xi$ at the unspecified time $\tau$ are equal to zero, no terminal condition is specified. According to the additional demand of the Pontryagin maximum principle [13] for an open-end OC problem the Hamiltonian Eq.(4) is equal to zero, $H(\xi^*,u^*,p^*)=0$, at each point of optimal trajectory (*). In such a formulation, the control $u$ has a physical sense of the changes in the rate constants. That makes the OC problem for dissipative kinetics as having much more sense in comparison to its artificial appearance when the classic mechanical problem was written in OC terms.

The problem of Eq.(3) could be rewritten as a variational problem in a similar way, as for mechanical case, when the control $u$ is evaluated from Eq.(2) and then substituted into Eq.(3):

$$S=\int_{t_0}^{\tau}(\Phi(\xi,\dot{\xi})+\Psi(\xi))dt\to\min,\ \xi(t_0)=\xi_0\,. \qquad (5)$$

Now, for an isolated thermodynamic system the Lagrange equations become

$$\frac{\partial^2\Phi}{\partial\dot{\xi}\partial\dot{\xi}}\ddot{\xi}+\frac{\partial^2\Phi}{\partial\xi\partial\dot{\xi}}\dot{\xi}+X^{\Psi}+X^{\Phi}=0\,. \qquad (6)$$

where $X^{\Psi}\equiv-\partial\Psi/\partial\xi$ are generalized forces related to free energy $\Psi$ and $X^{\Phi}\equiv-\partial\Phi/\partial\xi$ are generalized forces due to dependence of kinetic part $\Phi$ on extent coordinate. Since the Lagrange problem Eq.(5) is an open-end, so the transversality condition (TVC) applied:

$$\left(\Phi(\xi,\dot{\xi})+\Psi(\xi)-\dot{\xi}^T\partial\Phi/\partial\dot{\xi}\right)_{\xi^*,\tau}=H(\xi^*,\dot{\xi}^*)=0, \qquad (7)$$

where $\xi^*$ is the optimal trajectory, which has similar look to the additional Pontryagin maximum principle demand of the equality in similar open-end OC problem.

Applying the Legandre transform to thermodynamic Lagrangian from Eq.(5) we can obtain thermodynamic Hamiltonian

$$H(\xi,p)=p^T\dot{\xi}(p)-\Phi(\xi,\dot{\xi}(p))-\Psi(\xi), \qquad (8)$$

where $p=\partial\Lambda/\partial\dot{\xi}$ are thermodynamic momenta, and then the canonical system can be written. The TVC gives for the optimal trajectory $\xi^*$, that $H(\xi^*,p^*)=0$.

However when considering complete relaxation, the variational formulation is wholly consistent with an infinite-horizon formulation, which can be formulated as

$$S=\int_{0}^{\infty}(\Phi(\xi,\dot{\xi})+\Psi(\xi))dt\to\min,\ \xi(0)=\xi_0, \qquad (9a)$$

then the optimality conditions are similar to finite horizon problem with the exception of the transversality conditions, which in an infinite-horizon are

$$\lim_{t\to\infty}p(t)^T\xi(t)=0\,. \qquad (9b)$$

According to the Hamiltonian from Eq.(8) it is possible to write the Hamilton-Jacobi equation with respect to the MED principle. In a general form this equation could be written as

$$\partial S/\partial t+H(\xi,\partial S/\partial\xi,t)=0, \qquad (10)$$

where $S$ is thermodynamic action in an energetic representation and $H$ is thermodynamic Hamiltonian that explicitly depends on time. When the thermodynamic Hamiltonian is time-independent this equation can be written as

$$H(\xi,\partial S/\partial\xi)=E \qquad (11)$$

where $E$ is a constant ($E=0$ for optimal trajectory).

In classical linear case, when in the vicinity of the global equilibrium, the dissipation function $\Phi$ becomes a quadratic function [15] and can be written as [12], $2\Phi=\dot{\xi}^TR\dot{\xi}$ where $\dot{\xi}$ is the vector of the generalized displacement derivatives, $R$ is the positive definite matrix. The quadratic approximation of thermodynamic potential can be written as [15] $2\Psi=\xi^TL\xi$ where $\xi$ is the vector of generalized displacement, $L$ is the positive definite matrix. Then thermodynamic Lagrangian is

$$\Lambda=\Phi+\Psi=\frac{\dot{\xi}^TR\dot{\xi}}{2}+\frac{\xi^TL\xi}{2} \qquad (12)$$

and the Euler-Lagrange equations are similar to [12]

$$\ddot{\xi}=R^{-1}L\xi \qquad (13)$$

which can describe the exponential relaxation due to the positive definite scalar matrices $R$ and $L$. Second-order equation Eq. (13) describes both growing and decaying exponential branches and after applying boundary condition it selects the stable decaying branch. When $R$ and $L$ are scalar and symmetric then one can obtain [12]

$$J=\sqrt{R^{-1}L^{-1}}X\,, \qquad (14)$$

This equation shows that in terms of generalized thermodynamic fluxes $J$ and generalized thermodynamic forces $X$, the linear relations between the fluxes and forces have taken place in the vicinity of global equilibrium. The Hamiltonian correspondent to Eq.(12) is

$$H(\xi,p)=\frac{p^TR^{-1}p}{2}-\frac{\xi^TL\xi}{2} \qquad (15)$$

which gives the canonical system

$$\dot{\xi}=R^{-1}p,\dot{p}=L\xi \qquad (16)$$

and the Hamilton-Jacobi equation

$$(\partial S/\partial\xi)^TR^{-1}(\partial S/\partial\xi)-\xi^TL\xi=2E\,. \qquad (17)$$

## 3. Results

### 3.1 General linear optimal control case

Let as consider a general linear, relatively to the control variable $u$ case, when the relationship between rate vector $\dot{\xi}$ is given by a system of equations that are linear relative to vector $u$:

$$\dot{\xi}=F(u-k)+h \qquad (18)$$

where $\xi$, $u$ are $N$-dimensional vectors, $F$ is a $N*N$ matrix with coefficients $f_{ij}=f_{ij}(\xi_1,\ldots,\xi_N)$, $h=h(\xi)$. Let us take the cost function $\Lambda$ for the OC problem in a form, analogous to Eq.(3) when $\Phi(u-k)$ is quadratic:

$$\Lambda(\xi,u)=\frac{1}{2}(u-k)^TK(u-k)+\Psi(\xi) \qquad (19)$$

where $K$ is positive definite $N*N$ matrix. Following our approach we can reformulate the dynamic OC problem as a variational problem by substitution $(u-k)$ from Eq.(18) into Eq.(19), if matrix $F$ is a nonsingular around global equilibrium ($\xi=0$). Then we can write the Eq.(19) in terms of $\xi$ and $\dot{\xi}$, and obtain the variational Lagrangian

$$\Lambda(\xi,\dot{\xi}) = \frac{1}{2}(\dot{\xi}-h)^T R(\dot{\xi}-h) + \Psi(\xi) \quad (20)$$

where $R(\xi) = (F^{-1})^T K F^{-1}$ and matrix $F$ is a nonsingular. Then the Euler-Lagrange equations are

$$\ddot{\xi}^T (R^T + R) + (\dot{\xi}-h)^T \frac{\partial R^T}{\partial \xi}\dot{\xi} - \frac{\partial h^T}{\partial \xi}\dot{\xi}^T (R^T + R) + (\dot{\xi}-h)^T \frac{\partial (Rh)^T}{\partial \xi} + \frac{\partial h^T}{\partial \xi} R(\dot{\xi}-h) = 2\frac{\partial \Psi}{\partial \xi} \quad (21)$$

The Hamiltonian will be

$$H(\xi, p) = 2p^T Z p + p^T h - \Psi \quad (22)$$

where $Z \equiv (R^T + R)^{-1} R^T (R^T + R)^{-T}$ and canonical system is

$$\dot{\xi} = 2p^T (Z^T + Z) + h$$
$$\dot{p}^T = -2p^T (\partial Z/\partial \xi) p - p(\partial h/\partial \xi) + \partial \Psi/\partial \xi \quad (23)$$

The Hamilton-Jacobi equation is

$$2\left(\frac{\partial S}{\partial \xi}\right)^T Z \frac{\partial S}{\partial \xi} + \left(\frac{\partial S}{\partial \xi}\right)^T h - \Psi = E \cdot \quad (24)$$

### 3.2 Pure physical example – relaxation in an ideal RC circuit

In [11,12] we considered examples related to chemical kinetics. In contrast, let us illustrate a pure physical example, a so-called RC circuit. In this case, a capacitor ($C$ is its capacitance) is grounded by a resistor with resistance $R$. At $t_0$=0 the capacitor loaded to total charge $q_0$ and the discharging dissipative relaxation starts. By using our approach (see Eqs. (5)-(9), as well as Eqs. (18)-(23) ), this example can be written in one line. Taking into account Ohm's law, the constrained equation in the OC problem can be expressed as

$$\dot{q} = -u/R, \quad (25)$$

where $\dot{q}$ is a derivative of the charge $q$ (electric current, $I = \dot{q}$), $R$ is the resistance, and $u$ is a formal control. Now we need to build the cost function (Lagrangian) of the OC problem. Let us take the term corresponding to the potential (free energy of the capacitor) as

$$\Psi = \frac{q^2}{2C}$$

which is effectively the energy stored in the capacitor, $E_C = q^2/2C$. The penalty for formal regulation, corresponds to the dissipation function, we can take in a quadratic form $\Phi = Cu^2/2$. Then the OC Lagrangian following Eq.(19) will be

$$\Lambda_{RC} = \frac{q^2}{2C} + \frac{Cu^2}{2} \cdot$$

By substituting control $u$ from Eq. (25) into this OC Lagrangian, we can obtain the variational Lagrangian

$$\Lambda_{RC} = \frac{q^2}{2C} + \frac{R^2 C\dot{q}^2}{2}. \quad (26)$$

From this Lagrangian, we can obtain the Euler-Lagrange equation

$$\ddot{q} = \frac{1}{R^2 C^2} q \quad . \quad (27)$$

Using the boundary $q(0) = q_0$ or transversality condition $\left(q^2/2C - R^2 C\dot{q}^2/2\right)_{q^*,\tau} = 0$, or $(\dot{q}^*)^2 = (q^*)^2/R^2C^2$,

$$\dot{q}^* = \pm q^*/RC \quad (28)$$

which is a first-order equation, we then obtain

$$q^*(t) = q_0 \exp(-t/RC) \quad . \quad (29)$$

It is the well-known expression for electrical circuits.

By employing the Legendre transform $p = \partial \Lambda_{RC}/\partial \dot{q} = R^2 C\dot{q}$, we can build a variational Hamiltonian

$$H_{RC} = \frac{p^2}{2R^2C} - \frac{q^2}{2C}, \quad (30)$$

In fact, this equation, as well as equation (26), has an energy sense. It is well-known that the power dissipated by a resistor $R$ is $W = RI^2 = R\dot{q}^2$, so the first term of (30), as well as the second term of (26), can be written as $p^2/2R^2C = \dot{q}^2 R^2 C/2 = I^2 R^2 C/2$. The second term in Eq.(26) is $\Psi = q^2/2C = E_C$. Summarising, Eq.(30) can be rewritten as $H_{RC} = WRC/2 - E_C$ For the exponential RC discharge, $WRC/2$ equals to the instantaneous capacitor energy and hence the total energy that will be dissipated from the current time to infinity, $W$ changes continuously during discharge. One can note that the Hamiltonian $H_{RC}$ expression for the RC-circuit, Eq.(30), has an energy meaning and this meaning can be extended to any nonlinear dissipative process. Equation (28) can also be rewritten in terms of the generalised flux (electric current) $I_E = \dot{q}^*$ and generalised force $X_E$ as

$$I_E = X_E/R, \quad (31)$$

which shows the linear relation between electric flux and force. In fact, $X_E$ is the voltage. This example indicates the validity of Eqs.(18)-(23). The whole approach shown by Eqs.(3)-(11) and Eqs.(12)-(14) is not just limited to chemical thermodynamics but can be applied to physical thermodynamic processes, when the phenomenological kinetics can be written in terms of the extent from equilibrium. It follows from this example that, in a thermodynamic sense, the electric charge $q$ can, in some cases, be considered as a measure of the extent from global equilibrium.

More strict result of above formulated ideal RC circuit can be obtained from the infinite-horizon linear-quadratic (LQ) optimal control problem using Hamilton-Jacobi-Bellman(HJB)/algebraic Ricatti method, however in this way the result is more correct as trajectory Eq.(29) never riches $q = 0$. Formulation of infinite-horizon LQ OC problem can also start from Eq.(25) and $q(0) = q_0$. Then infinite-horizon cost can be written following Eq.(19) and Eq.(26) as $J(u) = \int_0^\infty (q^2/2C + Cu^2/2)dt \to \min_{u(\cdot)}$. This can be considered as a scalar LQR problem – to minimize quadratic running cost over an infinite horizon, subject to linear dynamics, driving charge $q$ to 0. If to define the value function $V(q) = \min_{u(\cdot)} J(u)$ according to above under-integral function, the problem is autonomous and $V$ solves the time-independent HJB equation $0 = \min_u [q^2/2C + Cu^2/2 + V(q)\dot{q}] = \min_u \cdot$ Using Eq.(25) to minimize over $u$ $\frac{\partial}{\partial u}\left[q^2/2C - V(q)u/R\right] = 0$, gives $u^*(q) = V(q)/RC$. Guessing $V(q) = Pq$, and substituting $u^*(q) = Pq/RC$ back into HJB equation, $0 = \frac{q^2}{2C} + \frac{C}{2}\left(\frac{Pq}{RC}\right)^2 - \frac{Pq}{R}\frac{Pq}{RC}$, or $\frac{q^2}{2C} - \frac{P^2q^2}{2R^2C} = 0$, then $P^2 = R^2$ and taking "+" root $P = R$, which requires for $V(q) \geq 0$ so $V(q) = \frac{Rq^2}{2}$ and then $u^*(q) = \frac{q}{C}$. From Eq.(25) for closed-loop dynamics $\dot{q} = -u^*/R = -q/RC$ and finally $q^*(t) = q_0 \exp(-t/RC)$ which coincides with Eq.(29) but reproduced in OC strict way. The transversality condition for infinite-horizon Eq.(9b) holds: $\lim_{t\to\infty} Rq_0^2 \exp(-2t/RC) = 0$. Finally $J^* = R\xi_0^2/2$ so the value function $V(\xi_0) = R\xi_0^2/2$.

### 3.3 Macroscopic cooperative and collective effects

Considering a one-dimensional example, in which Eq.(18) is written as $\dot{\xi} = f(\xi)u + h(\xi)$, we can write the Lagrangian for the pure variational problem according to Eq.(20) as:

$$\Lambda(\xi,\dot{\xi}) = \frac{r(\dot{\xi}-h)^2}{2f^2} + \Psi \qquad (32)$$

where $r$ is a constant, $r>0$. The Euler-Lagrange equation becomes

$$f\ddot{\xi} - f'_\xi(\dot{\xi}^2 - h^2) = fhh'_\xi + f^3\Psi'_\xi/r \,. \qquad (33)$$

According to Eq.(32) the Hamiltonian is

$$H(\xi,p) = p^2 f^2/2r + ph - \Psi \qquad (34)$$

and the canonical system

$$\begin{aligned} \dot{\xi} &= pf^2/r + h, \\ \dot{p} &= -p^2 ff'_\xi/r - ph'_\xi + \Psi'_\xi \end{aligned} \,. \qquad (35)$$

Finally, for logistical case, when $f(\xi)=1-\xi$, $\xi<1$, $h=0$, and quadratic potential $\Psi$, the Lagrangian is

$$\Lambda(\xi,\dot{\xi}) = r\frac{\dot{\xi}^2}{2(1-\xi)^2} + l\frac{\xi^2}{2} \,. \qquad (36)$$

Since Lagrangian is not explicitly dependent on time, and for finite-cost infinite-horizon solution relaxing to $\xi(t)\to 0$ $\dot{\xi}(t)\to 0$, the Beltrami first integral [16] is

$$\Lambda - \dot{\xi}\frac{\partial\Lambda}{\partial\dot{\xi}} = l\frac{\xi^2}{2} - r\frac{\dot{\xi}^2}{2(1-\xi)^2} = 0 \,. \qquad (37)$$

Then the stabilizing branch is

$$\dot{\xi}^* = -\sqrt{l/r}\,\xi^*(1-\xi^*) \qquad (38)$$

and

$$p^* = \sqrt{lr}\,\xi^*(1-\xi^*)$$

The transversality condition holds:

$$\lim_{t\to\infty} p(t)\xi(t) = \lim_{t\to\infty}\sqrt{lr}\frac{\xi^{*2}(t)}{1-\xi^*(t)} = 0 \,.$$

The Eq.(38) can be rewritten in terms of generalized thermodynamic flux and force as

$$J = \sqrt{r^{-1}l^{-1}}X^\Psi + \sqrt{r^{-1}l^{-3}}(X^\Psi)^2, \qquad (39)$$

which indicates that in the vicinity of global equilibrium ($\xi<<1$), when the generalized force is small, this expression coincides with the linear expression Eq.(14) when $X^\Psi = X$. Equation Eq.(39) can be written in terms of total force X as $J = \sqrt{l/rl}X/(l-X)^2$, then near-equilibrium expansion is

$$J = \sqrt{r^{-1}l^{-1}}X + 2\sqrt{r^{-1}l^{-1}}lX^2 + O(X^3) \,.$$

The canonical system obtained from corresponding Hamiltonian

$$H(\xi,p) = \frac{p^2(1-\xi)^2}{2r} - l\frac{\xi^2}{2} \qquad (40)$$

is

$$\begin{aligned} \dot{\xi} &= p(1-\xi)/r, \\ \dot{p} &= l\xi + p^2(1-\xi)/r \end{aligned} \,. \qquad (41)$$

The same form of equation for logistical kinetics, as Eq.(38), can be found using pure physical approach, based on proposed by Landau as a general theory for description of the second-order phase transitions and extended together with Ginzburg [14]. Usually the Landau free energy is taken as function of several parameters [15], including order-parameter. For our logistical cooperativity case, the Landau free energy can be simplified and taken in a form

$$\Psi_L(\xi) = l_L\frac{\xi^2(1-\xi)^2}{2}, \qquad (42)$$

where $\xi$ is state variable, $l_L>0$. Then using our approach we can construct the Lagrangian as the sum of quadratic dissipative function $\Phi_L = r_L\dot{\xi}^2/2$ and Landau-like free energy $\Psi_L$ from Eq.(42):

$$\Lambda_L(\xi,\dot{\xi}) = r_L\frac{\dot{\xi}^2}{2} + l_L\frac{\xi^2(1-\xi)^2}{2} \,. \qquad (43)$$

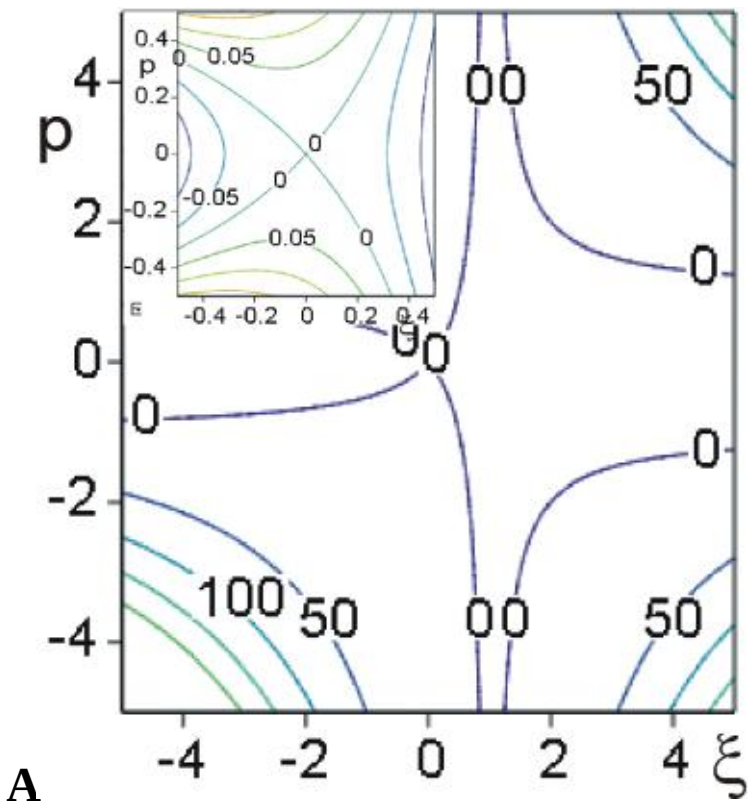

A

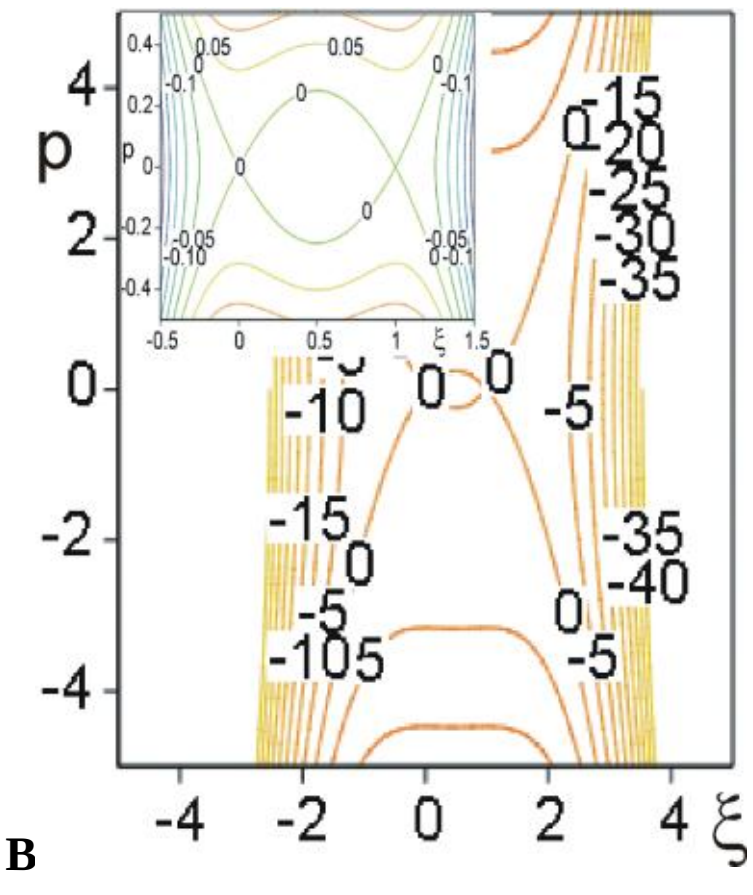

B

**Fig.1.** Logistical model, contour plot for the Hamiltonian: A, our approach, Eq.(40), $l=1$, $r=1$; and B, approach, based on employment of Landau-like free energy, Eq.(44), $l_L=1$, $r_L=1$.

Let us note that in both cases (our approach and based on the Landau free energy), the Euler-Lagrange equation looks identical, however the Hamiltonian

$$H_L(\xi,p) = p_L^2/2r_L - l_L\xi^2(1-\xi)^2/2 \qquad (44)$$

and canonical system

$$\dot{\xi} = p_L / r_L,$$
$$\dot{p}_L = l_L \xi(1-\xi)(1-2\xi) \qquad (45)$$

look differently in the Landau example.

Using the Beltrami first integral for Eq.(43) similarly to Eq.(37) we can obtain $\dot{\xi}* = -\sqrt{l_L / r_L}\,\xi * (1-\xi*)$, which coincides in form with Eq.(38).

**4. Discussion**

Fig.1 and Fig.2 compare two approaches, formulated in Eqs.(32)-(41) and Eqs.(42)-(45), for obtaining the logistic kinetics variational formulation. In our method, the thermodynamic momentum is $p* = -\sqrt{lr}\,\xi * /(1-\xi*)$, shown in Fig.1A as a phase trajectory $H=0$, while using Landau-like free energy, the momentum is $p_L* = -\sqrt{l_L r_L}\,\xi * (1-\xi*)$, phase trajectories are shown in Fig.1B, $H=0$ ; one can see significant differences in the contour plots. Fig.1 also shows the region $\xi < 0$ for illustration, which is in rather a non-physical region. These differences in the thermodynamic momenta follow from the differences in the Lagrangian definition, Eq.(36) and Eq.(43). In the case of the classical Landau free energy approach, the kinetic part of the Lagrangian (dissipative function) is quadratic, so the thermodynamic momentum in this case is linear to the velocity of the dissipative process with the coefficient $r_L$: $p_L = \partial\Lambda_L / \partial\dot{\xi} = r_L\dot{\xi}$. In our OC based interpretation, which takes free energy as an energetical penalty for being not in the equilibrium [11], the co-state variables or thermodynamic momenta are the energetical prices for the goal function Λ change due to an elementary change in the velocity of dissipative process. In such an OC sense, the energetical cost of dissipation is different in the case of Landau-like free energy and our approach. The phase trajectories in Fig.1 reflect this difference. Moreover, in our approach, the parameters in the Eqs.(2)-(3), Eq.(18) are the explicit rate constants and, therefore, the process is not symmetrical with respect to time; it is explicit asymmetry in phase plane regarding vertical axis, Fig.1A, see as well the insert in Fig.1A. In the approach based on Landau-like free energy, there is the symmetry between two states $\xi=0$ and $\xi=1$, Fig.1B. Taking as an example the second order phase transition, the difference between temperature $T$ and the critical temperature $T_C$, $r = r_0(T - T_C)$ plays the role of physical spanning parameter [15]. In this case, the phase transitions can go in a reverse direction because ideally we can change the temperature in any direction - in our example, from state $\xi=0$ to $\xi=1$. The contour plot in Fig.1B, therefore, produces a symmetrical characteristic, in contrast to the asymmetrical plot in our case, Fig.1A (compare as well as the inserts in Fig.1B and Fig.1A, correspondingly).

The differences shown in Fig.2 reflect additional specificity of our approach, where the generalized thermodynamic force X is a sum of $X^{\Psi}$ and $X^{\Phi}$, Eq.(6). Indeed, $X^{\Psi}$ is a summand that is due to the potential term Ψ dependence on the extent from equilibrium; this part of thermodynamic force is linear to the extent from equilibrium. Part of the generalized thermodynamic force $X^{\Phi}$ is due to the dependence of the kinetic part Φ on the extent coordinate, which characterizes the remoteness of the system from equilibrium. This particular dependence makes our approach specific: together these two parts, $X^{\Psi}$ and $X^{\Phi}$ create the general thermodynamic force X. Indeed, Fig.2 shows the generalized thermodynamic force X for logistic cooperativity as a function of the extent from equilibrium $\xi$. One can see a significant difference between the thermodynamic force in our approach, which is the sum of two parts $X_{Logistic} = X^{\Psi} + X^{\Phi}$, and those in the Landau-like free energy approach (as we mentioned, utilizing in classical description of nonlinearities mainly in the macroscopic phase transitions), $X_{Landau}$. However, in the region close to equilibrium, Fig.2B, the differences are not so significant and are vanishing when $\xi<<1$. In this region, the kinetics of obtaining equilibrium is close to exponential. It can also be seen from the $X_{Landau}$ plot in Fig.2A, that the generalized force looks anti-symmetrical relatively the point $\xi=0.5$. This graphically indicates that $\xi=0$ and $\xi=1$ are just symmetrical states of equilibrium, which also follows from Fig.1B, see as well as the insert in Fig.1B. While the state $\xi=1$ is an unstable state in our approach (as we have just one global equilibrium $\xi=0$), in fact, region $\xi<0$ is also in a non-physical region in Fig.1. An interesting impact in our approach is the effect of the "true" thermodynamic force $X^{\Psi}$ on the overall force $X_{Logistic}$. One can see from Fig.2A that it is a linear impact, while the impact of force $X^{\Phi}$, formally related to the kinetic part Φ, is significantly nonlinear. When $\xi<<1$ (near the equilibrium), the effect of this part of the generalized force is vanishingly small, Fig.2B.

A

B

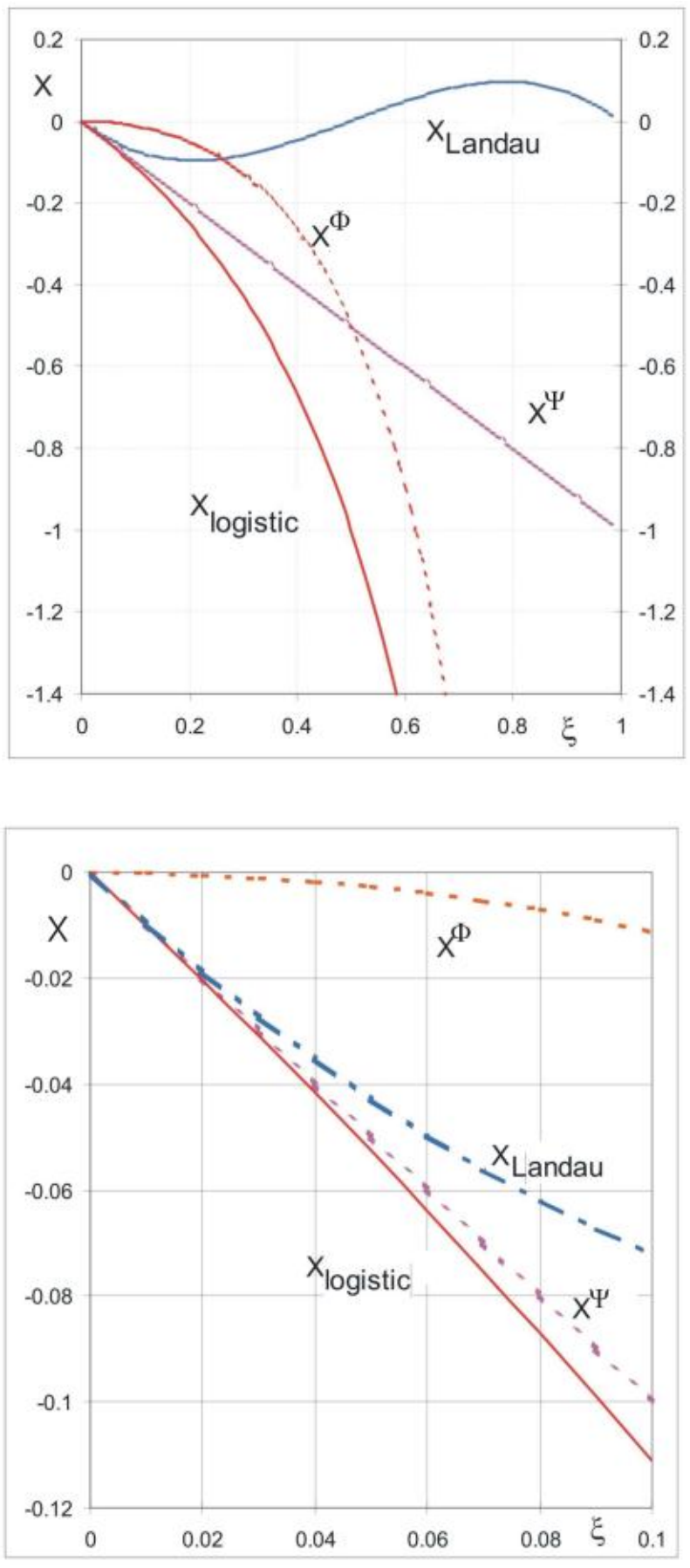

**Fig.2.** The generalised thermodynamic forces in logistical model for our approach, Eq.(40), $l$ =1, $r$ =1; ($X_{Logistic} = X^{\Psi}+X^{\Phi}$) and using the Landau potential Eq.(44), $l_L$ =1, $r_L$ =1 ($X_{Landau}$). **A**, $0<\xi<1$; **B**, $\xi<<1$ (in the vicinity of equilibrium).

Summarisingly, in the approach based on the Landau-like free energy, the nonlinearity is enclosed in the form of the thermodynamic potential Ψ, Eq.(42), while the dissipation function $\Phi$ is quadratic. In our approach, the nonlinearity is enclosed in the dissipative function dependence on the extent from equilibrium, while free energy is quadratic. Also, the nonlinearity is hidden (enclosed) in the manner of dissipation, while in the Landau-like formulation, the nonlinearity is only due to the potential part. This in fact reflects the essential difference between chemical-and-biological cooperativity, when our assumptions can easily go far inside the dissipative mechanisms, as in Eq.(2) and, from another side, between the collective phenomena in physics, when there is no indication of any organized character of dissipation and therefore the nonlinearity can be linked just to specific organization of the potential (free energy). Finally, consideration of the cooperative effects in chemical kinetics and the collective effects in physics illustrates the quite fundamental differences between these two phenomena. These differences are linked to the degree of irreversibility in these processes. Various phase transitions in macro-scale physics are reversible on principle. Temperature plays a crucial role as the spanning parameter (as it does in the Landau-like free energy); in some ideal cases by changing temperature it is possible to obtain any one of the stable states (equilibria). In contrast, in chemical, and especially, in biochemical kinetics, the irreversibility is more complex, hierarchical. In this case, changing temperature is not always sufficient to make the process irreversible and time plays a more explicit role.

Thus, the formal optimal control approach to nonlinear thermodynamics can be helpful in developing a variational approach based on the maximum energy dissipation principle. By applying this framework, we can transform the initial optimal control formulation to the variational approach. If we assume that the physical sense does not change, we can also apply the optimal control penalty interpretation to the variational approach. In this case, the thermodynamic Lagrangian can be interpreted as having an energetic penalty sense. Thermodynamic potential $\Psi$ can be treated as an energetic penalty for being removed from equilibrium. The dissipation function $\Phi$ can be interpreted as a penalty (energetical cost) for the capability to perform the dissipation, which also depends on the extent from equilibrium. The generalised thermodynamic momenta can be treated as the partial energetical loss due to the elementary changes in the dissipation velocities (generalised thermodynamic flows). The part $X^{\Psi}$ of the generalised thermodynamic forces X can be interpreted as the change in energetical loss $\Psi$ (free energy) due to elementary move towards the equilibrium. The part $X^{\Phi}$ of the generalised thermodynamic forces X – as the partial energetical loss related to an elementary move towards the equilibrium due to the existence/occurrence of dissipation, due to dependence of dissipation function on the state variable (extent from equilibrium).

When the constrained dynamic system in the optimal control problem is linear with regard to the control variables, the variational problem can be straightforwardly formulated. Furthermore, when the matrix of the dynamic constraint system has constant coefficients, and the thermodynamic potential has quadratic approximation (in the vicinity of equilibrium), the result of our approach coincides with linear non-equilibrium thermodynamics. The relaxation process in electric RC circuits illustrates good applicability of our approach to some relaxational physical processes. The cooperative/collective kinetics case, important for many applications, has been considered as a key nonlinear example, uniting physics and non-physics. The expression in terms of generalized thermodynamic forces and generalized thermodynamic fluxes has also been written for a logistical cooperative process, which appears as a quadratic approximation of general nonlinear relations.

## 5. Conclusion

In this Letter we have shown that the collective phenomena in physics and cooperative phenomena in chemistry/biology can be described in terms of proposed earlier variational framework [11,12]. Within the employment of the maximum energy dissipation principle, the cost-like functional can be chosen according to the optimal control initial formulation. It is suggested that using this approach, proposed variational outline can be extended for the non-equilibrium thermodynamics. We have here illustrated, with reference to the electrical RC circuit the applicability of approach to relaxational physical processes. The differences between the application of the approach to describing cooperative phenomena in chemical and biochemical kinetics, and the Landau free energy approach to collective phenomena in physics has been discussed within the proposed outline.